\documentclass[twocolumn,showpacs,showkeys,superscriptaddress]{revtex4}
\usepackage{graphicx,subfigure,mathrsfs}

\newcommand{\bastar}{\begin{eqnarray*}}
\newcommand{\eastar}{\end{eqnarray*}}
\newskip\humongous \humongous=0pt plus 1000pt minus 1000pt

\newif\ifdtup

\relax
\newcommand{\bea}{\begin{eqnarray}}
\newcommand{\eea}{\end{eqnarray}}
\newcommand{\nn}{\nonumber}
\newcommand{\pro}{\partial}
\newcommand{\oneg}{\dfrac{1}{g}}
\newcommand{\dfrac}{\displaystyle\frac}
\newcommand{\abc}{\alpha \beta \gamma}
\newcommand{\e}{\vec e}

\newcommand{\m}{\vec m}

\newcommand{\vl}{{\bf {l}}}
\newcommand{\vk}{{\bf {k}}}
\newcommand{\vp}{{\bf {p}}}
\newcommand{\X}{{\vec X}}

\newcommand{\A}{{\vec A}}
\newcommand{\B}{{\vec B}}

\newcommand{\vGm}{{\bf \Gamma}}

\newcommand{\vR}{{\bf R}}
\newcommand{\vZ}{{\bf Z}}

\newcommand{\vOm}{{\bf \Omega}}

\newcommand{\vtl}{\tilde{{\bf l}}}

\newcommand{\vtp}{\tilde{{\bf p}}}

\newcommand{\hn}{{\hat n}}

\newcommand{\hA}{{\hat A}}

\newcommand{\mn}{{\mu\nu}}

\begin{document}
\title {Knot Topology of Vacuum Space-Time and  \\
Vacuum Decomposition of Einstein's Theory}
\author{Y. M. Cho}
\email{ymcho@unist.ac.kr}
\affiliation{School of Electrical and Computer Engineering  \\ 
Ulsan National Institute of Science and Technology, Ulsan 689-805, Korea}
\affiliation{School of Physics and Astronomy,
Seoul National University, Seoul 151-742, Korea}
\author{Franklin H. Cho}
\affiliation{Department of Physics and Astronomy, University of Southern California,
Los Angeles, CA 90089, USA}
\begin{abstract}
~~~~~Viewing Einstein's theory as the gauge theory of Lorentz group, 
we construct the most general vacuum connections which have vanishing 
curvature tensor and show that the vacuum space-time can be classified 
by the knot topology $\pi_3(S^3)\simeq \pi_3(S^2)$ of $\pi_3(SO(3,1))$. 
With this we obtain the gauge independent vacuum decomposition of 
Einstein's theory to the vacuum and gauge covariant physical parts. 
We discuss the physical implications of our result in quantum gravity. 
\end{abstract}

\maketitle

An important goal in theoretical physics is to construct 
a decent quantum gravity. For this purpose, we must 
understand the structure of the classical vacuum space-time 
first. To do that it is not enough for us to solve the vacuum 
Einstein's equation,
\bea
&R_\mn-\dfrac12~R~g_\mn=0.
\eea
We must solve the true vacuum equation to obtain the most general vacuum 
space-time 
\bea
&R_{\mn\rho\sigma}=0.
\label{veq}
\eea 
{\it The purpose of this Letter is to construct the most general 
vacuum connection in terms of the tetrad, and to obtain the generally 
invariant vacuum decomposition of Einstein's theory to the vacuum part 
and the generally covariant physical part. Our result shows that 
the vacuum space-time can be classified by the knot topology 
$\pi_3(S^3)\simeq \pi_3(S^2)$.} 

It is well known that Einstein's theory can be viewed as a gauge 
theory \cite{uti,kib,prd76a}. In particular it can be viewed 
as a gauge theory of Lorentz group in which the gravitational 
connection and curvature tensor become the gauge potential and 
field strength of Lorentz group \cite{prd76b}. In this view 
the gauge invariance and the general invariance become synonymous, 
because each gauge transformation has the corresponding general 
coordinate transformation. Adopting this view and imposing 
the vacuum isometry, we construct all possible vacuum connections 
in Einstein's theory, and show that they are identical to 
the vacuum solutions of SU(2) gauge theory. 

This immediately tells that the vacuum in $R^4$ space-time has 
the knot topology identical to the $SU(2)$ vacuum. This 
raises the possibility of vacuum tunneling in Einstein's theory. 
Of course, ``the gravitational instantons" in Euclidian space-time 
classified by Euler-Poincare characteristics have been discussed 
before \cite{hawk,egu}. But they are discussed without any knowledge 
of topological classification of the vacuum in real space-time. 
Obviously we have to know the topological structure of the vacuum 
first to discuss the tunneling. Our result shows that it is the spin 
structure (i.e., the tetrad) of the flat space-time which describes 
the knot topology of the vacuum. 

Moreover, this allows the gauge invariant vacuum decomposition 
of the connection to the vacuum and gauge covariant physical parts. 
An important problem in Einstein's theory is to define the generally 
invariant momentum of gravitating particles which does not include 
the graviton. This has been thought to be impossible, because 
the covariant derivative always includes the connection. The vacuum 
decomposition not only makes this possible, but also allows us to 
obtain the vacuum decomposition of Einstein's theory itself.  

To discuss the structure of vacuum space-time we consider the $SU(2)$ 
gauge theory first. Let $\hn$ be an arbitrary gauge covariant unit 
triplet which selects the Abelian direction. Imposing the isometry 
\bea
D_\mu \hn=(\pro_\mu +g \vec A_\mu \times)~\hn=0,~~~~~(\hn^2=1)
\label{ccon}
\eea
we have the Abelian projection of $\vec A_\mu$ \cite{prd80,prl81}
\bea
&\vec A_\mu \rightarrow \hA_\mu= A_\mu \hn- \dfrac 1g \hn \times \pro_\mu \hn.
~~(A_\mu= \hn \cdot \A_\mu)
\label{rp}
\eea
An important feature of the ``Abelian" binding potential $\hA_\mu$ is 
that it retains the full $SU(2)$ gauge degrees of freedom, and is closed 
(transforms among itself) under the gauge transformation. Moreover, 
(\ref{rp}) allows us to have the Abelian decomposition \cite{prd80,prl81}
\bea
\A_\mu=\hA_\mu+\X_\mu,~~~~(\hn \cdot \vec X_\mu=0)
\eea 
where $\X_\mu$ is the gauge covariant valence potential. Notice that 
the decomposition is gauge independent. Once the Abelian direction 
is chosen the decomposition follows automatically, 
independent of the choice of a gauge.

Now, let $\hat n_i~(i=1,2,3)$ with $\hn_3=\hn$ be an orthonormal 
right-handed basis, and impose the vacuum isometry which assures 
the vanishing field strength 
\bea
\forall_i~~~D_\mu \hn_i =0.
\label{vcon}
\eea
Solving (\ref{vcon}) we obtain the most general
vacuum potential $\hat \Omega_\mu$ \cite{plb06}
\bea
&\hat \Omega_\mu = C_\mu \hn - \dfrac{1}{g} \hn 
\times \pro_\mu \hn = C_\mu^{~k}~\hn_k, \nn\\
&C_\mu^{~k} = \dfrac{1}{2g} \epsilon_{ij}^{~~k} 
(\hn_i \cdot \pro_\mu \hn_j),~~~C_\mu=-C_\mu^{~3}. 
\label{vac}
\eea
Notice that, although the vacuum is fixed by three isometries,
it is essentially fixed by $\hn$. This is because $\hn_1$ 
and $\hn_2$ are uniquely determined by
$\hn$, up to a $U(1)$ gauge transformation which leaves $\hn$
invariant. With
\bea
&\hn = \Bigg(\matrix{\sin{\alpha}\cos{\beta} \cr
\sin{\alpha}\sin{\beta} \cr \cos{\alpha}}\Bigg),
\label{n}
\eea
we have \cite{plb06}
\bea
&C_\mu^1= \oneg (\sin \gamma \pro_\mu \alpha
-\sin{\alpha} \cos \gamma \pro_\mu \beta), \nn\\
&C_\mu^2 = \oneg (\cos \gamma \pro_\mu \alpha
+\sin{\alpha} \sin \gamma \pro_\mu \beta), \nn\\
&C_\mu^3 = \oneg (\cos{\alpha} \pro_\mu \beta+\pro_\mu \gamma),
\eea
where the angle $\gamma$ represents the $U(1)$ angle which leaves
$\hn$ invariant. 

It is well known that the $SU(2)$ vacuum has the $\pi_3(S^3)$ 
topology \cite{thooft,bpst}. This topology is naturally inscribed 
in our vacuum (\ref{vac}). The vacuum quantum number which determines 
the wrapping number of the mapping from the compactified space $S^3$ 
to the $SU(2)$ space $S^3$ is given by the Chern-Simon index of 
$\hat \Omega_\mu$ \cite{plb06,thooft}
\bea
&n=-\dfrac{g^3}{96\pi^2} \int \epsilon_{0\abc} \epsilon_{ijk}
C_\alpha^i C_\beta^j C_\gamma^k d^3x. 
\label{nacsi}
\eea
But with the Hopf fibering this $\pi_3(S^3)$ can also be interpreted 
as $\pi_3(S^2)$ of the mapping from $S^3$ to the coset space
$SU(2)/U(1)$ defined by $\hn$ \cite{plb06}. Of course, 
the topologically distinct vacua become unstable under the quantum 
fluctuation because of the vacuum tunneling made by instantons. 
This leads us to introduce the $\theta$-vacuum 
\bea
|\theta \rangle= \dfrac{}{}\sum_n \exp (in\theta)~|n \rangle,
\eea
in non-Abelian gauge theory \cite{thooft,bpst}. 

Another important application of (\ref{vac}) is that it allows us to have 
the vacuum decomposition
\bea
&\vec A_\mu= \hat \Omega_\mu+\vec Z_\mu,
~~~\vec Z_\mu=(A_\mu+C_\mu)~\hn+ \vec X_\mu.
\label{vd}
\eea
Here again $\hat \Omega_\mu$ (just like $\hat A_\mu$) has the full 
gauge degrees of freedom, and $\vec Z_\mu$ transforms covariantly. 
Moreover, the decomposition is gauge independent. 

The vacuum decomposition allows us to define the gauge invariant canonical 
momentum of charged particles which does not include the gauge boson, which 
has been thought to be impossible \cite{jauch}. Indeed, in QCD (\ref{vd}) 
allows us to define the gauge invariant canonical momentum of quarks which 
does not include the gluons. This has important applications. In particular, 
this makes a gauge independent decomposition of nucleon momentum and spin 
to those of quarks and gluons possible \cite{nmsd}. 

Now, treating Einstein's theory as a gauge theory of 
Lorentz group, we can find the most general vacuum connection
imposing the vacuum isomstry. Let $J^{ab}~(a,b=0,1,2,3)$ 
be the six generators of Lorentz group which can be expressed by 
the rotation and boost generators $L_i$ and $K_i~(i=1,2,3)$. 
Let $\vp$ (or $p^{ab}$) be an adjoint representation of Lorentz 
group and $\vtp$ (or $\tilde p^{ab}=\epsilon_{abcd}~p^{cd}/2$) 
be its dual partner, and let $\m$ and $\e$ be the magnetic and electric 
components of $\vp$ which correspond to $3$-dimensional rotation 
and boost,
\bea
\vp=\left( \begin{array}{c} \m \\
\e \end{array} \right),
~~~\vtp=\left( \begin{array}{c} \e \\
-\m \end{array} \right),~~~\tilde {\vtp}=-\vp.
\eea
To proceed further let $\vGm_\mu$ (or $\Gamma_\mu^{~ab}$)
be the gauge potential of Lorentz group which describes the
spin connection, and $\vR_\mn$ (or $R_\mn^{~~ab}$)
be the curvature tensor
\bea
\vR_\mn=\pro_\mu \vGm_\nu-\pro_\nu \vGm_\mu
+\vGm_\mu \times \vGm_\nu.
\eea
Now, consider the following isometry
\bea
D_\mu \vp = (\pro_\mu + \vGm_\mu \times) ~\vp=0.
\label{ic}
\eea
This automatically assures
\bea
D_\mu \vtp =(\pro_\mu + \vGm_\mu \times) ~\vtp=0, 
\label{dic}
\eea
which tells that the isometry in Lorentz group 
always includes the dual partner. To verify this 
we decompose the gauge potential $\vGm_\mu$ into 
the 3-dimensional rotation and boost parts
$\A_\mu$ and $\B_\mu$, 
\bea
\vGm_\mu= \left( \begin{array}{c} \A_\mu \\
\B_\mu \end{array} \right).
\eea
With this both (\ref{ic}) and (\ref{dic}) can be written 
as \cite{grg1}
\bea
&D_\mu \m= \B_\mu \times \e,
~~~~D_\mu \e= -\B_\mu \times \m, \nn\\
&D_\mu = \pro_\mu + \A_\mu \times.
\label{ic1}
\eea
This confirms that (\ref{ic}) and (\ref{dic}) are actually identical
to each other, which tells that the isometry in Lorentz
group always comes in pairs.

Let $\vl_i$ and $\vk_i$ be an orthonormal basis of the adjoint 
representation of Lorentz group which describe the rotation and 
boost. And let $\hn_i$ be the orthonormal triplets of the $SU(2)$ 
subgroup of Lorentz group,
\bea
&\vl_i= \left( \begin{array}{c} \hn_i \\
0 \end{array} \right),
~~~\vk_i= \left( \begin{array}{c} 0 \\
\hn_i  \end{array} \right)= -\vtl_i.
\label{lbasis}
\eea
Now, it must be clear that the most general vacuum connection 
which guarantees $\vR_\mn=0$ is described by the following 
vacuum isometry
\bea
&\forall_i~~~D_\mu \vl_i =0,
~~~D_\mu \vk_i = -D_\mu \vtl_i= 0.
\label{vic}
\eea
Actually the vacuum needs only one of them, because 
they are dual to each other.

To obtain the most general vacuum connection we have to solve (\ref{vic}).
To do that, notice that in $3$-dimensional notation (\ref{vic})  
is written as 
\bea
&\forall_i~~~D_\mu \hn_i= \B_\mu \times \hn_i,
~~~D_\mu \hn_i= -\B_\mu \times \hn_i.
\label{vic1}
\eea
This has the unique solution
\bea
&\A_\mu=\hat \Omega_\mu,~~~\B_\mu=0.
\label{gvac1}
\eea
where $\hat \Omega_\mu$ is precisly the vacuum potential 
(\ref{vac}) of the $SU(2)$ subgroup. So we have the most general
vacuum connection $\vOm_\mu$ which yields vanishing curvature tensor 
\bea
\vGm_\mu=\vOm_\mu=\left( \begin{array}{c} \hat \Omega_\mu \\
0 \end{array} \right).
\label{gvac2}
\eea
This shows that the vacuum connection of Einstein's theory is
given by the vacuum potential of $SU(2)$ gauge theory.

Obviously this tells that the vacuum space-time has exactly the same
structure as the vacuum of $SU(2)$ gauge theory. In particular, 
this tells that the vacuum (in $R^4$ space-time) can be classified 
by the knot topology $\pi_3(S^3) \simeq \pi_3(S^2)$. This may not be 
surprising, considering the fact $\pi_3(SO(3,1))\simeq \pi_3(SU(2))$.

At this point, one may wonder if one could describe this vacuum 
topology in terms of the metric. To answer this, consider a flat 
space-time which has an $R^4$ topology. Choosing a proper coordinate 
basis $\pro_\mu$, we can always transform the flat metric and 
torsionless flat Levi-Civita connection to aquire the trivial form,  
\bea
&g_\mn=\eta_\mn,~~~~~\Gamma_{\mn}^{~~\alpha}=0.
\label{fc1}
\eea 
Now, introduce an orthonormal (i.e., tetrad) basis $e_a$ by
\bea
&e_a=\hn_a^{~\alpha} \pro_\alpha,~~~\pro_\mu=\hn_\mu^{~a} e_a,
~~~(\hn_\mu^{~a} \hn_{a \nu}=\eta_\mn)  \nn\\
&e_0=\pro_0,~~e_i=\hn_i^{~\alpha}~\pro_\alpha,  
~~~(\hn_0^{~\alpha}=\delta_0^{~\alpha},~\hn_i^{~0}=0) \nn\\
&[e_a,~e_b]=f_{ab}^{~~c} e_c,  \nn\\
&f_{ab}^{~~c}=(\hn_a^{~\mu}\pro_\mu \hn_b^{~\nu}
-\hn_b^{~\mu}\pro_\mu \hn_a^{~\nu}) \hn_\nu^{~c}.
\label{t}
\eea
Using the identity
\bea
{\mathscr D}_\mu e_\nu^{~a}\equiv \pro_\mu e_\nu^{~a}
-\Gamma_\mn^{~~\alpha} e_\alpha^{~a}+\Gamma_{\mu b}^{~~a} e_\nu^{~b}=0,
\label{id}
\eea
we can easily show that the flat connection (\ref{fc1}) acquires 
the following form in the tetrad basis
\bea
&\Gamma_\mu^{~ab}=\dfrac{\eta_{\alpha \beta}}2 \big(\hn_a^{~\alpha}
~\pro_\mu \hn_b^{~\beta}-\hn_b^{~\alpha}
~\pro_\mu \hn_a^{~\beta} \big).
\label{gvac3}
\eea
This is precisely the vacuum connection (\ref{gvac2}). Moreover 
(with $\hn_\mu^{~a} \hn_{a \nu}=\eta_\mn$) we can show that this 
is nothing but the flat connection (\ref{fc1}) expressed in 
the tetrad basis,
\bea
&\Gamma_\mu^{~ab}=\dfrac12 \hn_\mu^{~c}(f_c^{~ab}-f_c^{~ba}-f^{ab}_{~~c})
=\Omega_\mu^{~ab}.
\eea
This tells that it is the spin structure of the flat space-time, 
the non-trivial configuration of tetrad, which describes the vacuum 
topology. In particular, this shows that metric and torsion 
have no role in the vacuum topology. 

It has generally been assumed that the tetrad is no more fundamental 
than the metric. Moreover, ever since Feynman tried to quantize the gravity 
with the metric, the metric has been treated as the quantum field of 
gravity \cite{feyn}. But our result shows that the tetrad is more 
fundamental. In fact, there is an unshakable evidence for this: The graviton 
which couples to spinors in Feynman diagrams is the tetrad, not the metric. 
So, when spinors are present, we have no choice but to treat the tetrad as 
the fundametal field of gravity. 

Clearly our result raises the possibility of the vacuum tunneling in Einstein's 
theory.  Candidates of the ``gravitational instanton" which have finite 
Euclidian action have been discussed by many authors \cite{hawk,egu}. 
But they have been proposed without any reference to the above vacuum 
topology. Whether any of these, or any unknown gravito-instantons, 
can actually demonstrate the tunneling is an interesting question 
worth further study \cite{cqg11}. 

Assuming the tunneling, we can certainly introduce the $\theta$-vacuum 
in gravity. The question is whether this $\theta$-vacuum could be 
the physical vacuum in quantum gravity. Of course, in gauge theory 
the answer is no. But in Einstein's theory the $\theta$-vacuum 
could play an important role, because a gravito-instanton (if exists) 
could have a vanishing action and thus have the maximum tunneling 
probability. This could make the $\theta$-vacuum much more important 
in quantum gravity. 

But perhaps a more important application of (\ref{gvac2}) is that it 
provides us the Lorentz invariant vacuum decomposition of an arbitrary 
connection to the vacuum and gauge covariant physical parts by
\bea
&\vGm_\mu= \vOm_\mu + \vZ_\mu,  \nn\\
&\vR_\mn=\bar D_\mu \vZ_\nu-\bar D_\nu \vZ_\mu
+\vZ_\mu \times \vZ_\nu, 
\label{vdec1}
\eea
where $\bar D_\mu= \pro_\mu+ \vOm_\mu \times$. To understand the meaning 
of this, let $\mbox{\boldmath$\alpha$}$ be an infinitesimal gauge parameter. 
Now, under the gauge transformation 
\bea
&\delta \vGm_\mu= D_\mu \mbox{\boldmath$\alpha$},
~~~\delta \vl_i=-\mbox{\boldmath$\alpha$} \times \vl_i,
~~~\delta \vk_i=-\mbox{\boldmath$\alpha$} \times \vk_i,  \nn
\eea 
we have 
\bea
&\delta \vOm_\mu= {\bar D}_\mu \mbox{\boldmath$\alpha$},
~~~\delta \vZ_\mu= -\mbox{\boldmath$\alpha$} \times \vZ_\mu.
\eea
This tells that $\vOm_\mu$ retains the full Lorentz gauge degrees of 
freedom and forms a closed connection space by itself. Moreover, 
$\vZ_\mu$ transforms covariantly and thus can be interpreted to 
represents the physical part of the connection. Most importantly, 
the decomposition is gauge independent. Once $\vl_i$ and $\vk_i$ 
are chosen, the decomposition follows independent of the gauge. 

Certainly we can have a similar generally invariant vacuum decomposition 
in the coordinate basis. Using (\ref{id}) we can transform $\vOm_\mu$ 
and $\vZ_\mu$ to the coordinate basis, 
\bea
&\Omega_\mn^{~~\alpha}= \Omega_\mu^{~ab} e_{a \nu} e_b^{~\alpha}
+e_a^{~\alpha}\pro_\mu e_\nu^{~a},  \nn\\
&Z_\mn^{~~\alpha}=Z_\mu^{~ab} e_{a \nu} e_b^{~\alpha}.
\eea
Notice that $\Omega_\mn^{~~\alpha}$ (just like $\Omega_\mu^{~ab}$) transforms 
exactly as a connection under the general coordinate transformation, 
and forms its own closed connection space. In contrast, 
$Z_\mn^{~~\alpha}$ (just like $Z_\mu^{~ab}$) transforms covariantly. 
With this we have the vacuum decomposition of an arbitrary connection 
to the vacuum part  and the generally covariant physical part 
in the coordinate basis \cite{grg2}
\bea
&\Gamma_\mn^{~~\alpha}=\Omega_\mn^{~~\alpha}+Z_\mn^{~~\alpha}, \nn\\
&R_{\mn \rho}^{~~~~\sigma}= \bar \nabla_\mu Z_{\nu\rho}^{~~\sigma}
-\bar \nabla_\nu Z_{\mu\rho}^{~~\sigma} 
+Z_{\mu \alpha}^{~~\sigma} Z_{\nu \rho}^{~~\alpha} \nn\\   
&-Z_{\nu \alpha}^{~~\sigma} Z_{\mu \rho}^{~~\alpha},  
\label{vdec2}
\eea
where $\bar \nabla_\mu$ is the generally covariant derivative made of
the vacuum connection. Clearly the decomposition is independent of choice 
of the general coordinates, because it holds in any coordinate basis.

A fundamental problem in gauge theory is to obtain a gauge independent 
canonical momentum of a charged particle which does not include 
the gauge boson \cite{jauch}. The vacuum decomposition (\ref{vd}) 
makes this possible \cite{nmsd}. Our vacuum decomposition in Einstein's 
theory plays an equally important role. To see this consider a gravitational 
binary system made of two particles, and try to decompose the momentum 
and spin of this system to those of the constituent particles and 
pure gravity in a generally invariant way. Without the vacuum decomposition 
this is impossible. But now this becomes possible, because  
(\ref{vdec2}) allows us to define the generally invariant canonical 
momentum of the particles which does not include the graviton \cite{grg2}.  

{\it In general the vacuum decomposition allows us to have the generally 
invariant energy-momentum and angular momentum decompositions of 
a composite system to those of pure gravitational and non-gravitational 
parts. More importantly, this provides us the generally invariant 
vacuum decomposition of Einstein's theory itself, and allows us to reformulate 
the theory essentially in terms of the generally covariant physical
quantities.} This will have far reaching consequences. In particular,
this could play an important role in quantum gravity.  

The detailed discussions on these and related 
subjects will be presented in a separate paper \cite{grg1,grg2}.
 
{\bf ACKNOWLEDGEMENT}

The work is supported in part by National Research Foundation
of Korea (Grant 2010-002-1564) and by Ulsan National Institute of Science 
and Technology.

\end{document}